\documentclass[twocolumn,showpacs,preprintnumbers,amsmath,amssymb]{revtex4-2}
\usepackage{graphicx}
\usepackage{dcolumn}
\usepackage{bm}
\begin{document}
\title{Melting phase diagram of the two-dimensional electron solid in a perpendicular magnetic field}
\author{M.~Yu.~Melnikov, D.~G. Smirnov, and A.~A. Shashkin}
\affiliation{Institute of Solid State Physics, Chernogolovka, Moscow District 142432, Russia}
\author{S.-H. Huang and C.~W. Liu}
\affiliation{Department of Electrical Engineering and Graduate Institute of Electronics Engineering, National Taiwan University, Taipei 106, Taiwan}
\author{S.~V. Kravchenko}
\affiliation{Physics Department, Northeastern University, Boston, MA 02115, USA}
\begin{abstract}
We determine the thermal melting phase diagram of the electron solid by measuring the onset density of the double-threshold voltage-current curves in the insulating state in the ultra-high mobility two-dimensional electron system in SiGe/Si/SiGe quantum wells at different temperatures in perpendicular magnetic fields. The introduction of a magnetic field leads to a considerable increase in the onset density, which is strongest at the lowest temperatures and decreases with increasing temperature, and the melting temperature increases with magnetic field. These data are in agreement with theoretical predictions. Unexpectedly, the melting temperature as a function of electron density in a fixed magnetic field reveals a plateau, which can be interpreted in terms of resonance modes in the crystallites.
\end{abstract}
\maketitle
\setlength{\parskip}{0pt}

The ground state of a two-dimensional (2D) electron system in the strongly interacting limit at low electron densities is expected to be a Wigner crystal \cite{wigner1934on,chaplik1972possible}. The crystallization is expected when the interaction parameter determined by the ratio of the Coulomb and Fermi energies, $r_{\text{s}}=g_{\text{v}}/(\pi n_{\text{s}})^{1/2}a_{\text{B}}$, reaches about 35 \cite{tanatar1989ground} (here $g_{\text{v}}=2$ is the valley degeneracy, $n_{\text{s}}$ is the electron density, and $a_{\text{B}}$ is the effective Bohr radius in semiconductor). The application of a perpendicular magnetic field is predicted to promote the formation of the crystalline state, because the magnetic field decreases the amplitude of the zero-point oscillations of the electrons at the lattice sites, thereby stabilizing the lattice (see, \textit{e.g.}, Refs.~\cite{lozovik1975crystallization,ulinich1979phase,fukuyama1975two,eguiluz1983two}). It has been predicted that the Wigner crystal in a single-valley 2D electron system should form at Landau filling factors below approximately 0.15 \cite{lam1984liquid,levesque1984crystallization}. An increase in the electron density of the crystallization, at least in zero magnetic field, is expected due to the presence of a residual disorder (see, \textit{e.g}., Ref.~\cite{chui1995impurity}). The phase diagram of the 2D electron system at finite temperatures in magnetic fields has been calculated (see, \textit{e.g.},  Refs.~\cite{ulinich1979phase,platzman1992the,chang1983melting,chui1991finite,fukuyama1979two,lozovik1980two,tsukada1977two}). There have been reports of the observation of a magnetically-induced Wigner crystal on semiconductor surfaces, based on the observation of reentrant insulating phases and single-threshold current-voltage ($I$-$V$) characteristics in the insulating phases around the integer and fractional quantum Hall states, attributed to the depinning of the Wigner crystal (see, \textit{e.g.},
Refs.~\cite{andrei1988observation,williams1991conduction,diorio1992reentrant,qiu2012connecting,knighton2014reentrant,deng2016commensurability,qiu2018new,knighton2018evidence,hossain2022anisotropic,falson2022competing,madathil2023moving,madathil2024signatures,seiler2025signatures}). However, alternative traditional interpretations of the data were discussed, such as percolation \cite{jiang1991magnetotransport,dolgopolov1992metal,shashkin1994insulating} or Efros-Shklovskii variable range hopping in strong electric fields \cite{marianer1992effective}. Reports of the measurements of the thermal melting phase diagrams of the Wigner solid have been published in Refs.~\cite{chen2006melting,rosales2021competition, knighton2018evidence, drichko2016melting, ma2020thermal}. It is worth noting that much confusion was introduced by the fact that many authors chose to interpret their data in terms of a Wigner crystal, ignoring mundane interpretations. It is also worth noting that the Wigner crystal states in question, which are expected to spontaneously form without any periodic potential, differ from the generalized Wigner crystal states on a lattice background discussed in Refs.~\cite{li2021imaging,tsui2024direct}.

Recently, two-threshold $V$-$I$ characteristics have been observed at low electron densities in high-mobility silicon metal-oxide-semiconductor field-effect transistors (Si MOSFETs) and ultra-clean SiGe/Si/SiGe quantum wells in magnetic fields that reveal the signature of a quantum Wigner solid and exclude simple interpretations in terms of percolation or overheating \cite{brussarski2018transport,melnikov2024collective}. The $V$-$I$ characteristics are similar to those observed for the collective depinning of the vortex lattice in type-II superconductors \cite{blatter1994vortices}, provided the voltage and current axes are interchanged. The obtained results can be described by a phenomenological theory of the collective depinning of elastic structures, which corresponds to thermally activated transport accompanied by a peak in generated broadband current noise between the dynamic and static thresholds, and to the solid sliding with friction as a whole over a pinning barrier above the static threshold. The predicted magnetic-field stabilization of the quantum electron solid has been confirmed in measurements of the double-threshold $V$-$I$ characteristics in SiGe/Si/SiGe quantum wells at lowest temperatures \cite{melnikov2025stabilization}. It is natural to associate the onset of the double-threshold $V$-$I$ curves with the melting point of the electron solid.

The high uniformity of samples is crucial for observing the double-threshold voltage-current characteristics, and the findings should be universal for other 2D electron systems \cite{melnikov2024collective}. In high-mobility Si MOSFETs, the disorder has a short-range character caused by the residual point scatterers at the interface and the low-temperature mobility reaches approximately 3 m$^2$/Vs, whereas in a class of the purest accessible semiconductor heterostructures (including electron and hole GaAs/AlGaAs heterostructures, ZnO/MgZnO heterostructures, AlAs quantum wells, SiGe/Si/SiGe quantum wells), the long-range disorder potential due to background impurities is present and the carrier mobility is higher by orders of magnitude. However, it is well known that mobility is a crude characteristic of the disorder potential. In particular, the long-range harmonics of disorder potential are crucial at the lowest carrier densities in the insulating regime but become screened as the carrier density increases into the metallic regime \cite{ando1982electronic}. As inferred from both the level and character of the disorder potential, Si MOSFETs and unprecedentedly high-mobility heterostructures belong to different classes of electron systems. Similar results obtained in Si MOSFETs and SiGe/Si/SiGe quantum wells show the generality of the effect for different classes of electron systems.

In this paper, we determine the thermal melting phase diagram of the electron solid by measuring the onset density of the double-threshold voltage-current curves in the insulating state in the ultra-high mobility two-valley two-dimensional electron system in SiGe/Si/SiGe quantum wells at different temperatures in perpendicular magnetic fields. The introduction of a magnetic field increases the onset density, which is strongest at the lowest temperatures and decreases with increasing temperature. The melting temperature increases with increasing magnetic field. These data are consistent with theoretical predictions. Unexpectedly, the melting temperature as a function of electron density in a fixed magnetic field reveals a plateau, which can be interpreted in terms of resonance modes in the crystallites. The expected behavior of the plateau level as a function of magnetic field or electron density is consistent with the experimental results.

The samples used are ultra-high mobility SiGe/Si/SiGe quantum wells similar to those described in Refs.~\cite{melnikov2015ultra,melnikov2017unusual,melnikov2024triple}. The low-temperature electron mobility reaches $\approx200$~m$^2$/Vs at a density $\approx10^{11}$~cm$^{-2}$ in these samples. The approximately 150~\AA\ wide silicon (001) quantum well is sandwiched between Si$_{0.8}$Ge$_{0.2}$ potential barriers. The samples were patterned in Hall-bar shapes with the distance between the potential probes of 100~$\mu$m and a width of 50~$\mu$m using standard photolithography. The triple-top-gate design included the main Hall-bar gate, the contact gate, and the depleting gate, separated by SiO as an insulator; the use of the contact gate dramatically reduced the contact resistances and the use of the depleting gate suppressed the shunting conduction channel between the contacts outside the Hall bar, which manifested itself at the lowest electron densities in the insulating regime (for more details, see Ref.~\cite{melnikov2024triple}). No intentional doping was used, and the electron density was controlled by applying a positive dc voltage to the gate relative to the contacts. We applied saturating infrared illumination to the samples for several minutes, after which the contact quality improved, and the electron mobility increased, as had been found empirically \cite{melnikov2015ultra,melnikov2017unusual}. The resistance of the contacts measured below 10~kOhm. Measurements were conducted in an Oxford TLM-400 dilution refrigerator equipped with a magnet. The dc voltage was applied between the source and the nearest potential probe over a distance of 25~$\mu$m. The current was measured by a current-voltage converter connected to a digital voltmeter. The voltage-current curves exhibited slight asymmetry upon reversing the voltage. We plotted the negative part versus the absolute value of the voltage for ease of representation. The electron density was determined by analyzing Shubnikov-de~Haas oscillations in the metallic regime using a standard four-terminal lock-in technique; measurements were conducted within a frequency range of 1--11 Hz, while maintaining currents of 0.04--4 nA within the linear response regime. Experiments were performed on three samples, and the obtained results were similar.

\begin{figure}
\scalebox{.9}{\includegraphics[width=\columnwidth]{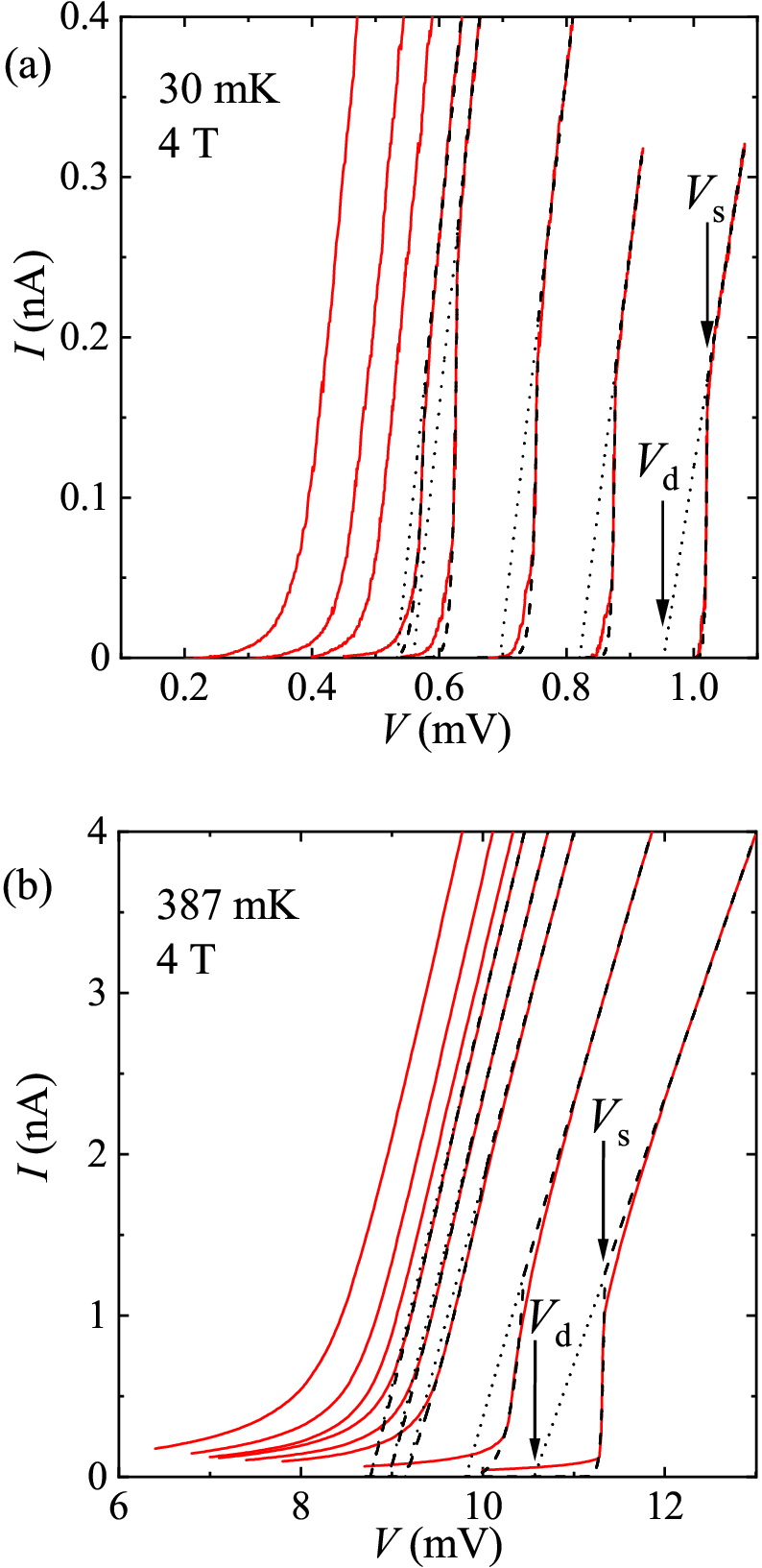}}
\caption{(a) Voltage-current characteristics in a magnetic field of 4~T at a temperature of 30~mK at electron densities (in units of $10^{10}$~cm$^{-2}$, left to right): 2.94, 2.85, 2.81, 2.78, 2.76, 2.67, 2.58, 2.49. (b) Voltage-current characteristics in a magnetic field of 4~T at a temperature of 387~mK at electron densities (in units of $10^{10}$~cm$^{-2}$, left to right): 0.99, 0.94, 0.90, 0.88, 0.85, 0.81, 0.72, 0.63. Also shown are the dynamic threshold $V_{\text d}$ obtained by the extrapolation (dotted line) of the linear part of the $V$-$I$ curves to zero current and the static threshold $V_{\text s}$. The dashed lines are fits to the data, see text.}
\label{fig1}
\end{figure}

\begin{figure}
\scalebox{1}{\includegraphics[width=\columnwidth]{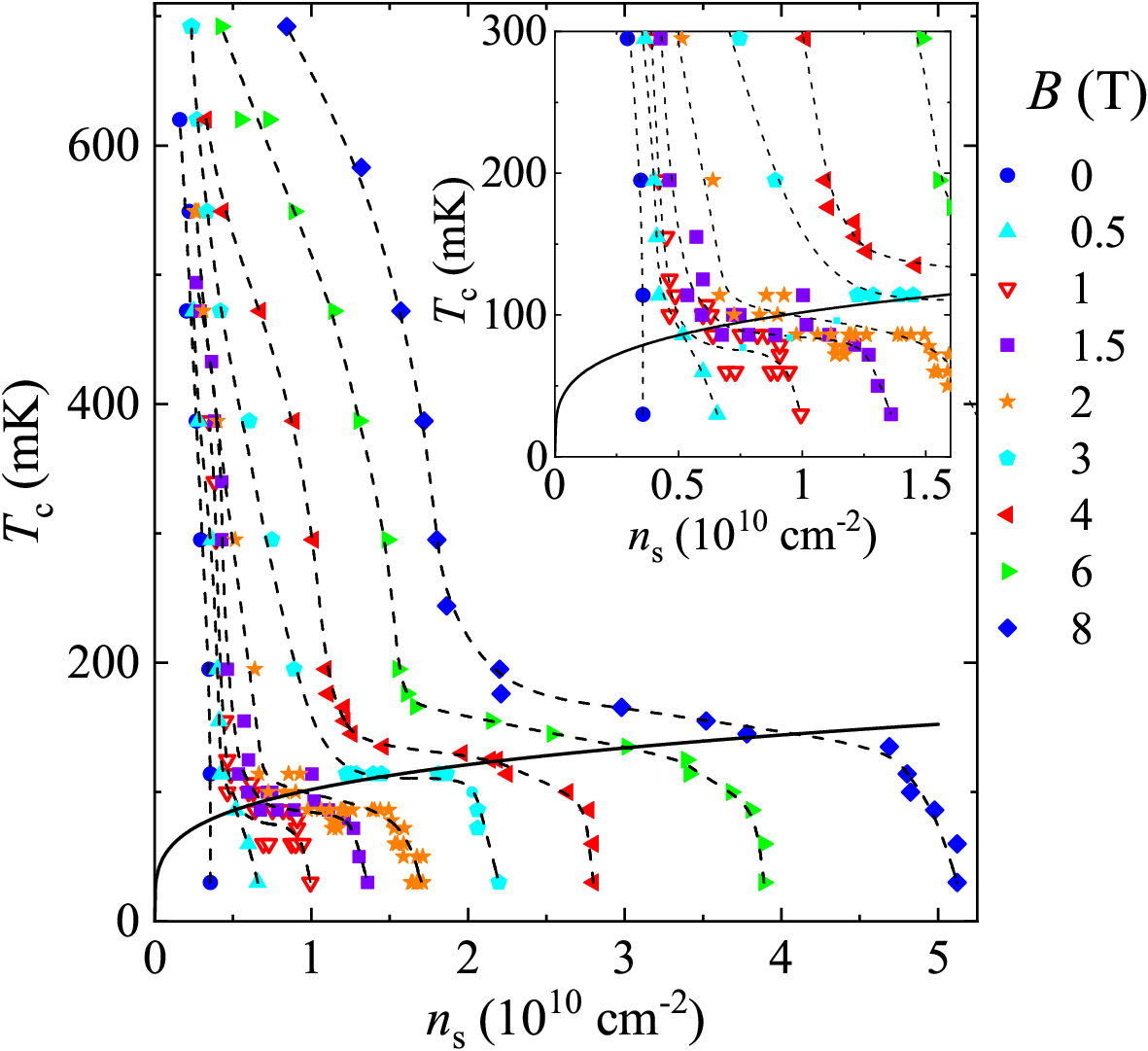}}
\caption{The melting temperature as a function of electron density in different magnetic fields. The dashed lines are guides to the eye. The expected dependence of the temperature on the plateau $T_{\text p}(n_{\text{s}})\propto n_{\text{s}}^{1/4}$ with a fitting parameter $l$ is shown by the solid line. The inset shows the low-$n_{\text{s}}$ region on an expanded scale.}
\label{fig2}
\end{figure}

In Fig.~\ref{fig1}(a, b), we show typical voltage-current characteristics at different electron densities in the low-density insulating state in a magnetic field $B=4$~T at temperatures $T=30$ and 387~mK; note that the insulating state in $B=0$ corresponds to the interaction parameter $r_{\text{s}}$ that exceeds 20 in this electron system. Two-threshold voltage-current curves are observed at electron densities below some onset density. As the applied voltage is increased, the current stays near zero until the first threshold voltage is reached. Then, the current sharply increases until the second threshold voltage is reached, above which the slope of the voltage-current curves is reduced significantly, and the behavior becomes linear, although it is not ohmic. This observation is consistent with the results obtained previously \cite{brussarski2018transport,melnikov2024collective,melnikov2025stabilization,melnikov2025inequivalence}. The double-threshold $V$-$I$ curves shift to higher threshold voltages as the electron density is reduced and one advances into the low-density insulating state. The peak of broadband current noise between the two threshold voltages (not shown here) is observed in the absence or presence of the magnetic field. In this paper, we focus on the range of electron densities near the onset of the two-threshold $V$-$I$ characteristics, the onset density being taken at the midpoint between the neighboring single-threshold and double-threshold $V$-$I$ curves in the figure.

A phenomenological theory of the collective depinning of elastic structures was adapted for an electron solid in Ref.~\cite{brussarski2018transport}. As the applied voltage increases, the depinning of the electron solid is signaled by the appearance of a current. Between the dynamic ($V_{\text d}$) and static ($V_{\text s}$) thresholds, the collective pinning of the electron solid occurs, and the transport is thermally activated: $I=\sigma_0\,(V-V_{\rm d})\,\exp[-U_{\rm c}(1-V/V_{\rm s})/k_{\rm B}T]$, where $U_{\rm c}$ is the maximal activation energy of the pinning centers, $\sigma_0$ is a coefficient, and $V_{\rm d}$ corresponds to the pinning force. When the applied voltage exceeds the static threshold, the electron solid slides with friction over a pinning barrier, as determined by the balance of the electric, pinning, and friction forces, which results in linear $V$-$I$ characteristics: $I=\sigma_0\,(V-V_{\rm d})$. The corresponding fits, shown by the dashed lines in Fig.~\ref{fig1}(a, b), describe the experimental two-threshold $V$-$I$ characteristics well.

Associating the onset density, $n^*$, of the two-threshold $V$-$I$ curves with the melting point of an electron solid, similar to the low-temperature limit, one can plot the thermal melting phase diagram of the electron solid at elevated temperatures. The corresponding melting temperature $T_{\text c}$ as a function of electron density is shown in Fig.~\ref{fig2} for different magnetic fields. The introduction of a magnetic field promotes the formation of the electron solid in the studied range of temperatures, which manifests itself in a considerable increase of the value $n^*$. The effect of the magnetic field on $n^*$ is strongest at the lowest temperatures and decreases with increasing temperature. In turn, the temperature $T_{\text c}$ increases with increasing magnetic field, which is consistent with the magnetic-field stabilization of the electron solid. Remarkably, there arises a plateau in the dependence of $T_{\text c}$ on electron density in magnetic fields $B\gtrsim 1$~T. The plateau expands and shifts to higher temperatures as the magnetic field is increased.

The observed increase of the onset density $n^*$ for the double-threshold $V$-$I$ characteristics with increasing magnetic field indicates the stabilization of the electron solid in perpendicular magnetic fields. At the lowest temperatures, this is consistent \cite{melnikov2025stabilization} with the theoretical prediction that the application of a perpendicular magnetic field should promote the formation of the Wigner solid by decreasing the amplitude of the zero-point vibrations of the electrons at the lattice sites \cite{lozovik1975crystallization,ulinich1979phase,fukuyama1975two,eguiluz1983two}. According to Ref.~\cite{ulinich1979phase}, at not too low temperatures, the calculated mean square amplitude of the vibrations of the electron lattice includes a phonon-related contribution that is determined by temperature. As the temperature is increased, the phonon-related contribution should lead to weakening the dependence of the amplitude of the electron lattice vibrations on the magnetic field. This prediction is consistent with the observed reduction of the effect of the magnetic field on the onset density $n^*$ with increasing temperature in Fig.~\ref{fig2}. Also, the observed increase of the temperature $T_{\text c}$ with increasing magnetic field is concurrent with the predictions of Refs.~\cite{ulinich1979phase,lozovik1980two}.

The occurrence of a plateau in the dependence of the melting temperature on electron density in high magnetic fields is quite unexpected. Assuming the existence of crystallites in the electron solid, the observed plateau can be interpreted in terms of resonance modes in the crystallites. When the resonance modes become excited with increasing temperature, the amplitude of the electron lattice vibrations sharply increases, giving rise, in accordance with the Lindemann criterion, to a drop in the onset density. This drop in $n^*$ manifests itself as a plateau in the dependence $T_{\text c}(n_{\text{s}})$. One can estimate the expected behavior of the plateau level with the magnetic field or the electron density. The temperature $T_{\text p}$ on the plateau corresponds to the resonance frequency $f$ of the transverse mode in a crystallite and can be evaluated at $T_{\text p}=hf/k_{\text B}\sim h(\mu/mn_{\text{s}})^{1/2}/2lk_{\text B}\sim h(e^2/\varepsilon ma)^{1/2}/2lk_{\text B}$. Here the shear modulus $\mu$ in high magnetic fields is close to that in $B=0$ \cite{maki1983static,wang1989effective,esfarjani1991shear,ferconi1993dynamical} given by $\mu\sim e^2/\varepsilon a^3$, $m$ is the electron mass, $h$ is the Planck constant, $k_{\text B}$ is the Boltzmann constant, $e$ is the electron charge, $\varepsilon$ is the dielectric constant, $a$ is the interelectron distance, and $l$ is the crystallite size. The resulting dependence $T_{\text p}(n_{\text{s}})\propto n_{\text{s}}^{1/4}$ with a fitting parameter $l\sim 4$~$\mu$m is depicted by the solid line in Fig.~\ref{fig2}. This intercepts all the dependences $T_{\text c}(n_{\text{s}})$ in high magnetic fields within the plateaus, which is consistent with the experiment and favors the above interpretation. The fact that the plateau occurs in high magnetic fields may be related to the improved uniformity of the electron system observed in perpendicular $B$ \cite{melnikov2025stabilization}, in which case a more pronounced crystallite structure can be expected. Note that the crystallite size evaluation is consistent with the estimate made in line with Refs.~\cite{williams1991conduction, madathil2024signatures}, disregarding numerical factors: $l\sim (e/\varepsilon E_{\text{th}})^{1/2}\sim 2$~$\mu$m, where $E_{\text{th}}$ is the threshold electric field estimated by the threshold voltage $\sim 1$~mV near the observed plateau (Fig.~\ref{fig1}(a) and Fig.~\ref{fig2}).

In summary, we have determined the thermal melting phase diagram of the electron solid by measuring the onset density of the double-threshold voltage-current curves in the insulating state in the ultra-high mobility 2D electron system in SiGe/Si/SiGe quantum wells at different temperatures in perpendicular magnetic fields. Applying a magnetic field causes a considerable increase in the onset density, strongest at the lowest temperatures but decreasing with increasing temperature; the melting temperature also increases with magnetic field, in agreement with theoretical predictions. Unexpectedly, the melting temperature as a function of electron density in a fixed magnetic field reveals a plateau, which can be interpreted in terms of resonance modes in the crystallites. Further theoretical consideration of the effects is needed.

This investigation was supported by RSF (project no. 25-22-00708). The SiGe/Si/SiGe wafers were grown under support by the Ministry of Science and Technology, Taiwan (Project No.\ NSTC 114-2634-F-A49-001).


%

\end{document}